\documentclass[conference]{IEEEtran}
\IEEEoverridecommandlockouts
\usepackage{cite}
\usepackage{amsmath,amssymb,amsfonts}
\usepackage{algorithmic}
\usepackage{algorithm}
\usepackage{graphicx}
\usepackage{textcomp}
\usepackage{xcolor}
\usepackage{booktabs}
\usepackage{array}
\usepackage{url}
\usepackage{hyperref}
\usepackage{tikz}
\usetikzlibrary{arrows.meta,positioning,shapes,shadows}

\def\BibTeX{{\rm B\kern-.05em{\sc i\kern-.025em b}\kern-.08em
    T\kern-.1667em\lower.7ex\hbox{E}\kern-.125emX}}

\IEEEpubid{\begin{minipage}{\textwidth}\ \\[3\baselineskip]  
    \centering 979-8-3195-0703-7/26/\$31.00\copyright~2026~IEEE
\end{minipage}}

\begin{document}

\title{Closed-Loop Dynamic Validator Node Scaling in Private Substrate Blockchains Using Takagi-Sugeno Fuzzy Inference}

\author{\IEEEauthorblockN{Thandile Nododile, Ayinde M. Usman, Clement N. Nyirenda}	
\IEEEauthorblockA{Department of Computer Science}
\IEEEauthorblockA{University of the Western Cape, South Africa}
\IEEEauthorblockA{0009-0000-1386-2425, 0000-0003-1926-3508, 0000-0002-4181-0478}}

\maketitle

\begin{abstract}
Private blockchain networks operate with fixed node configurations that cannot adapt to changing workload conditions. When too many nodes serve a light workload, resources are wasted; when too few nodes face heavy demand, block production slows and finalisation degrades. The right number of validator nodes is hard to determine, as it depends on multiple overlapping factors that shift over time. This paper presents a Takagi-Sugeno (TS) fuzzy inference system that reads live blockchain parameters, namely block production time, block size, and active node count, and outputs a continuous efficiency score alongside a scaling recommendation: \textit{Scale Up}, \textit{Maintain}, or \textit{Scale Down}. The controller uses triangular membership functions across three linguistic variables, evaluated through a complete 27-rule base with product t-norm aggregation. A central methodological contribution is an empirical recalibration of the membership functions, anchoring the linguistic terms to the observed operating range of the testbed rather than to theoretical extremes. The system is evaluated on a 10-node Substrate blockchain network storing real smart water meter data hashes from the Queensland Government open data portal. Statistical analysis across validator configurations of 4, 7, and 10 active nodes confirms that the controller produces distinct operational profiles that correctly reflect each configuration's provisioning state. In closed-loop experiments, the controller autonomously adjusts validator participation in both directions, activating validators under rising load and removing them under over-provisioning, and converges to the same stable equilibrium from both directions. Compared against three threshold-based baselines, it exhibits substantially fewer scaling oscillations while maintaining comparable block production times. The results demonstrate that TS fuzzy inference can support autonomous validator management in private blockchain deployments, with stable scaling behaviour that threshold approaches cannot match. 
\end{abstract}


\section{Introduction}
\label{sec:introduction}

Blockchain technology provides a distributed, tamper-resistant ledger architecture suitable for applications requiring data integrity and auditability \cite{nakamoto2008bitcoin}. Devices in Internet of Things (IoT) environments generate large volumes of data that often lack inherent security. Blockchain offers secure, tamper-resistant storage for such sensor data across domains including healthcare, supply chain, and utility monitoring \cite{dai2019blockchain, reyna2018blockchain}. Private blockchain networks, where validator participation is restricted to a known set of nodes, are particularly suited to enterprise IoT deployments that require auditability without the overhead of permissionless consensus \cite{wood2024substrate}. Deploying these networks for IoT data management, however, introduces a tension: validator configurations are typically fixed, while IoT workloads vary with environmental conditions, time-of-day patterns, and device availability \cite{Suryavansh2021}. A configuration sized for peak load wastes computational resources during quiet periods, while one sized for average load cannot maintain acceptable block production times during demand spikes \cite{jabbar2023enhancing}.

The challenge of adaptive node scaling is that the decision boundaries are not crisp. Network conditions exist on a continuum, and the appropriate scaling response depends on multiple interacting factors. When block production time is rising while block size is moderate and a subset of the available validators is already active, simple threshold rules cannot determine the correct action. Threshold-based controllers, common in distributed systems, react sharply at fixed cut-offs and tend to oscillate when input variables hover near the boundary. Fuzzy logic \cite{zadeh1965fuzzy} addresses this through degrees of membership, allowing multiple inputs to contribute proportionally to a decision. The Takagi-Sugeno (TS) fuzzy inference system \cite{takagi1985sugeno} is particularly suited to control applications: it produces crisp numerical outputs through weighted combinations of constant or linear functions and has been applied in networked control systems with communication constraints \cite{zhang2021ts}.

Prior work has established blockchain architectures for IoT data integrity \cite{nododile2023swm}, integrated Substrate with the InterPlanetary File System (IPFS) for off-chain storage \cite{nododile2024ipfs}, and demonstrated hybrid blockchain-IPFS solutions on smart water meter data \cite{nododile2025hybrid}. These works addressed security and storage efficiency but used static validator configurations. Dynamic scaling of validator participation, driven by the live state of the chain, remains an open problem.

This paper introduces a TS fuzzy inference system for closed-loop dynamic node scaling in private Substrate blockchains. The controller monitors block production time, block size, and active node count in real time, and triggers validator activation through verified peer discovery when scaling is recommended. A methodological contribution is the empirical recalibration of membership function parameters, anchored to the observed operating range of the testbed rather than to theoretical extremes. The controller is compared against three threshold-based baselines on identical workloads, with statistical tests characterising the trade-off between scaling stability and reactivity.

The experimental evaluation focuses on three aspects of system behavior. First, it examines whether the controller’s recommendations align with the underlying blockchain operational state, such that it suggests Scale Up when the system is under-provisioned, Maintain when operating near target conditions, and Scale Down when over-provisioned. Second, it assesses whether scaling actions drive block production time toward the effective slot duration target corresponding to the current validator set size, indicating closed-loop convergence. Third, it evaluates the balance between stability and responsiveness by comparing whether the TS controller exhibits fewer scaling oscillations than threshold-based controllers while maintaining similar convergence performance.

This paper's contributions include: (i) a closed-loop dynamic node scaling system in which the TS controller triggers actual validator activation with peer-discovery verification; (ii) an empirical recalibration of fuzzy membership functions with a generalisable anchoring rule and worked examples; (iii) a closed-loop evaluation that demonstrates both scaling directions, with validator activation under rising load and deactivation under over-provisioning, each converging to the same equilibrium; and (iv) a comparative evaluation against three threshold-based baselines with Welch's t-tests and Cohen's \textit{d} effect sizes.

\section{Related Work}
\label{sec:related}

\subsection{Blockchain Scalability and Node Management}
\label{subsec:blockchain-scaling}

Blockchain scalability is a well-known challenge \cite{zhou2020solutions, xie2019survey}, with the tension between decentralisation, security, and throughput \cite{buterin2021trilemma} driving research into dynamic node management. Substrate provides a modular framework where consensus is separated into block authoring through the Authority Round (AURA) algorithm and finality through GHOST-based Recursive Ancestor Deriving Prefix Agreement (GRANDPA) \cite{wood2024substrate}. These mechanisms operate on static, pre-configured rules; when transaction rates change, fixed validator configurations cannot adapt \cite{jabbar2023enhancing}.

Protocol-level approaches such as sharding \cite{luu2016elastico, kokoris2018omniledger} partition the network to process transactions in parallel but introduce cross-shard coordination complexity. He et al. \cite{he2022scalable} proposed a dynamic node selection approach for IoT blockchains using Graph Convolutional Networks, reducing latency but introducing computational overhead that may be prohibitive for resource-constrained edge deployments. Lightweight, adaptive mechanisms are still needed, that scale blockchain infrastructure without protocol modifications or heavy computation.

\subsection{Fuzzy Logic for Distributed Resource Management}
\label{subsec:fuzzy-distributed}

Fuzzy set theory \cite{zadeh1965fuzzy} extends classical set theory by allowing partial membership, providing a mathematical framework for handling uncertainty in decision-making. The TS fuzzy inference system \cite{takagi1985sugeno} uses crisp mathematical functions as rule consequents, which approximates nonlinear systems efficiently. Kaur and Kaur \cite{kaur2012mamdani} found Sugeno inference more computationally efficient than Mamdani for real-time control, and Zhao and Bose \cite{zhao2019membership} showed that triangular membership functions balance simplicity with control accuracy. TS models have been applied to networked control systems with communication constraints \cite{zhang2021ts}.

Fuzzy logic has been applied to resource allocation in cloud and edge contexts \cite{sharma2020cloud}. Closer to blockchain, Alouache et al. \cite{k2022vehicular} applied fuzzy logic to real-time reconfiguration in blockchain-based vehicular networks, and Omondi et al. \cite{ahmad2026fuzzy} applied fuzzy evaluation to smart contract access control. These works target vehicular networks and access control policies rather than infrastructure-level validator node management in private blockchain deployments.

\subsection{Blockchain-IoT Data Storage}
\label{subsec:blockchain-iot}

The growth of IoT deployments generates large volumes of sensor data requiring secure, distributed storage \cite{reyna2018blockchain}. Blockchain offers data integrity through its append-only structure but faces storage and throughput limitations when processing continuous streams \cite{dai2019blockchain}. Combining blockchain with IPFS addresses storage constraints by keeping large data objects off-chain while recording content hashes on-chain \cite{ali2017iot}, with applications across healthcare, supply chain, and industrial IoT \cite{maftei2023iot}. A line of prior work has developed blockchain architectures for IoT data integrity, including a Substrate-based mechanism for smart water meter data \cite{nododile2023swm}, Substrate-IPFS integration for distributed storage \cite{nododile2024ipfs}, and a hybrid blockchain-IPFS solution \cite{nododile2025hybrid}. These implementations achieved data security and storage efficiency but used static node configurations. The literature establishes that blockchain-IoT integration is well developed and that fuzzy logic effectively handles resource allocation under uncertainty. However, to the best of the authors' knowledge, there is no existing work that applies fuzzy inference to closed-loop blockchain node orchestration. This paper addresses that gap.

\section{System Design}
\label{sec:system}

This section describes the architecture of the proposed adaptive node scaling system and the design of the TS fuzzy inference engine that produces its scaling recommendations. It develops the first two contributions stated in Section \ref{sec:introduction}: the closed-loop scaling architecture with peer-discovery verification (Section \ref{subsec:architecture}, contribution i) and the empirical recalibration of the membership functions with a generalisable anchoring rule (Section \ref{subsec:mf-calibration}, contribution ii). Algorithm \ref{alg:scaling} summarises the closed-loop procedure.

\subsection{Architecture Overview}
\label{subsec:architecture}

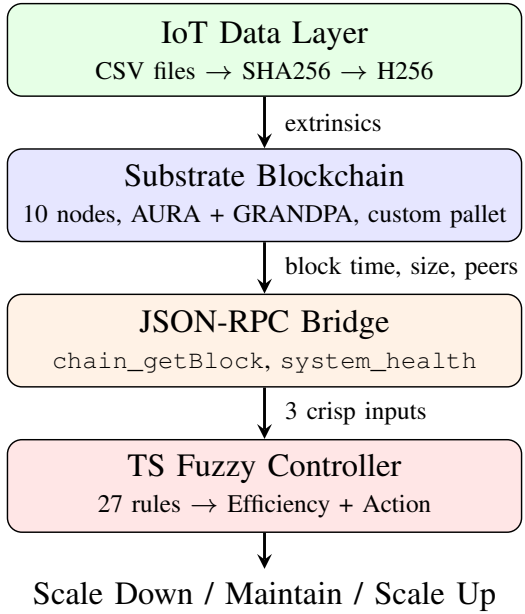
\begin{figure}[t]
\centering
\resizebox{0.80\columnwidth}{!}{%
\begin{tikzpicture}[
    node distance=0.5cm,
    box/.style={draw, rounded corners, minimum width=5.0cm, minimum height=0.9cm, align=center, font=\small},
    arrow/.style={->, >=stealth, thick},
    label/.style={font=\scriptsize, midway, right=2pt}
]
\node[box, fill=green!10] (iot) {IoT Data Layer\\{\scriptsize CSV files $\rightarrow$ SHA256 $\rightarrow$ H256}};
\node[box, fill=blue!10, below=of iot] (chain) {Substrate Blockchain\\{\scriptsize 10 nodes, AURA + GRANDPA, custom pallet}};
\node[box, fill=orange!10, below=of chain] (rpc) {JSON-RPC Bridge\\{\scriptsize \texttt{chain\_getBlock}, \texttt{system\_health}}};
\node[box, fill=red!10, below=of rpc] (fuzzy) {TS Fuzzy Controller\\{\scriptsize 27 rules $\to$ Efficiency + Action}};
\node[below=0.35cm of fuzzy, font=\small] (output) {Scale Down / Maintain / Scale Up};
\draw[arrow] (iot) -- node[label] {extrinsics} (chain);
\draw[arrow] (chain) -- node[label] {block time, size, peers} (rpc);
\draw[arrow] (rpc) -- node[label] {3 crisp inputs} (fuzzy);
\draw[arrow] (fuzzy) -- (output);
\end{tikzpicture}
}%
\caption{System architecture. IoT data files are hashed and submitted to the blockchain. The JSON-RPC bridge extracts live parameters, which the TS controller evaluates to produce scaling recommendations.}
\label{fig:architecture}
\end{figure}

The system comprises four functional layers, illustrated in Fig. \ref{fig:architecture}. At the top is the IoT data ingestion pipeline that submits Secure Hash Algorithm 256-bit (SHA-256) hashes of sensor data files to the blockchain through a custom runtime pallet. Below it sits the private Substrate blockchain network, running AURA consensus for block production and GRANDPA for deterministic finalisation. The third layer is a JavaScript Object Notation Remote Procedure Call (JSON-RPC) monitoring bridge that extracts live network parameters from the chain. At the bottom sits the TS fuzzy inference engine, which processes the bridge's three crisp inputs (block production time, block size, active node count) and outputs scaling recommendations. When a recommendation crosses the appropriate decision threshold for a sustained reading, the controller activates an additional validator, maintains the current configuration, or deactivates an excess validator.

The blockchain stores IoT data through a custom pallet that accepts SHA256 hashes of data files via a \texttt{storeDataHash(H256)} extrinsic. Raw data remains off-chain; only verification hashes are recorded on the ledger. Higher submission rates increase the number of extrinsics per block, which is one of the fuzzy controller's monitored inputs.

\begin{algorithm}[t]
\caption{Closed-Loop Validator Node Scaling}
\label{alg:scaling}
\begin{algorithmic}[1]
\STATE $n \gets$ initial active validators; $cooldown \gets 0$
\LOOP
    \STATE wait 5 s; read block time $bt$, block size $bs$, node count $n$
    \STATE fuzzify $(bt, bs, n)$ over all linguistic terms
    \STATE compute firing strengths $w_i$ (product t-norm) for the 27 rules
    \STATE $a \gets \left(\textstyle\sum_i w_i c_i\right) / \left(\textstyle\sum_i w_i\right)$ \COMMENT{action value}
    \IF{$cooldown > 0$}
        \STATE $cooldown \gets cooldown - 5$ \COMMENT{suppress decisions}
    \ELSIF{$a \geq 0.7$ \AND $n < 10$}
        \STATE activate validator; verify peer discovery
        \STATE $n \gets n + 1$;\; $cooldown \gets 30$ \COMMENT{Scale Up}
    \ELSIF{$a < 0.3$ \AND $n > 4$}
        \STATE deactivate highest-numbered validator (SIGTERM)
        \STATE $n \gets n - 1$;\; $cooldown \gets 30$ \COMMENT{Scale Down}
    \ELSE
        \STATE maintain configuration \COMMENT{Maintain}
    \ENDIF
\ENDLOOP
\end{algorithmic}
\end{algorithm}

\subsection{Input Variables and Membership Function Calibration}
\label{subsec:mf-calibration}

\begin{figure}[t]
    \centerline{\includegraphics[width=0.38\textwidth]{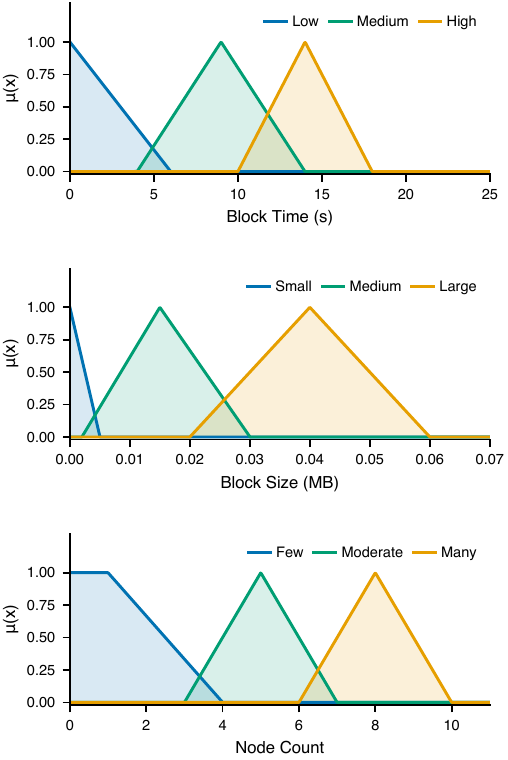}}
    \caption{Triangular membership functions for the three input variables after empirical calibration. Overlapping regions between adjacent terms enable smooth transitions in the controller's output. The Block Time and Node Count partitions are anchored to the observed operating range of the testbed.}
    \label{fig:membership}
\end{figure}

Three linguistic variables capture the blockchain network's operational state: block production time, block size, and active node count. Each variable is partitioned into three triangular membership functions (MFs), defined by three points $[a, b, c]$ representing the left foot, peak, and right foot. Fig. \ref{fig:membership} shows the calibrated MFs used in this work.

The operating range of these variables depends on the underlying Substrate configuration and the data pallet used. A controller calibrated for one chain may produce flat or saturated outputs on another. The MF parameters are therefore anchored to the observed operating range rather than to theoretical extremes:

\begin{itemize}
    \item \textbf{Block Time:} MFs are anchored to the empirically observed range. Block times across configurations of 4, 7, and 10 active validators span 6.0 to 14.4 s, so the High triangle peaks at 14 s (the observed worst case) rather than at the theoretical upper bound from the slot duration formula.
    \item \textbf{Block Size:} MFs are anchored to the empirical 50th and 95th percentiles of observed block sizes for the hash-storage pallet, mapping linguistic terms to the actual data distribution on the deployed chain.
    \item \textbf{Node Count:} MFs use a fixed partition with the Many term peaking at 8 active validators (the realistic ``fully scaled'' point) rather than at the chain ceiling of 10.
\end{itemize}

Table \ref{tab:mf-params} lists the calibrated MF parameters used in this work. Table \ref{tab:mf-examples} shows worked examples of how the same anchoring rule transfers to chains with different slot durations.

\begin{table}[t]
\caption{Calibrated Membership Function Parameters}
\label{tab:mf-params}
\centering
\begin{tabular}{lll}
\toprule
\textbf{Variable} & \textbf{Term} & \textbf{Parameters [a, b, c]} \\
\midrule
Block Time (s)    & Low      & [0, 0, 6] \\
                  & Medium   & [4, 9, 14] \\
                  & High     & [10, 14, 18] \\
\midrule
Block Size (MB)   & Small    & [0, 0, 0.005] \\
                  & Medium   & [0.002, 0.015, 0.030] \\
                  & Large    & [0.020, 0.040, 0.060] \\
\midrule
Node Count        & Few      & [1, 1, 4] \\
                  & Moderate & [3, 5, 7] \\
                  & Many     & [6, 8, 10] \\
\bottomrule
\end{tabular}
\end{table}

\begin{table}[t]
\caption{Worked Examples: Anchoring Rule Across Deployment Configurations}
\label{tab:mf-examples}
\centering
\footnotesize
\begin{tabular}{lccc}
\toprule
\textbf{Deployment} & \textbf{Slot} & \textbf{Active/Auth.} & \textbf{Effective Slot} \\
\midrule
This work (4-of-10 baseline)  & 6 s  & 4/10  & 15.0 s \\
This work (full topology)     & 6 s  & 10/10 & 6.0 s \\
Fast Substrate (3 s slot)     & 3 s  & -    & 3.0 s \\
Slow Substrate (12 s slot)    & 12 s & -    & 12.0 s \\
\bottomrule
\end{tabular}
\end{table}

\subsection{Output Variables}
\label{subsec:outputs}

Each rule produces two zero-order TS constants: an Efficiency Score from 0 to 100 reporting how effectively the network utilises its resources, and an Action Value from 0 to 1 that maps to a scaling recommendation through threshold partitioning. An action value below 0.3 recommends \textit{Scale Down} (decrease validator count by one), a value from 0.3 up to 0.7 recommends \textit{Maintain} (no change), and a value of 0.7 or above recommends \textit{Scale Up} (increase validator count by one).

\subsection{Rule Base}
\label{subsec:rule-base}

The rule base consists of 27 rules covering every combination of three input variables with three terms each ($3 \times 3 \times 3 = 27$). This completeness guarantees that no input state produces a default fallback response; every operating condition maps to a specific efficiency assessment and action recommendation. The rules encode domain knowledge about the relationship between blockchain operating conditions and appropriate scaling responses. When block time is Low and blocks are Small, the network has excess capacity and the action consequent favours \textit{Scale Down} (e.g.\ R1: Low BT, Small BS, Few NC $\to$ Eff = 90, Act = 0.15, where BT denotes block time, BS block size, NC node count, Eff the efficiency score, and Act the action value). When block time is High and blocks are Large with Few active nodes, the network is severely under-provisioned and the consequent favours \textit{Scale Up} (e.g.\ R25: High BT, Large BS, Few NC $\to$ Eff = 20, Act = 0.90). Balanced conditions favour \textit{Maintain} (e.g.\ R14: Medium BT, Medium BS, Moderate NC $\to$ Eff = 70, Act = 0.50). Efficiency consequents span [20, 90] and action consequents span [0.15, 0.90], with the full rule set generated by interpolating between these anchor points along each input axis.

\subsection{TS Fuzzy Inference}
\label{subsec:ts-inference}

The controller employs a zero-order TS fuzzy inference system \cite{nyirenda2011fuzzy}. Unlike Mamdani systems \cite{nyirenda2008self, kaur2012mamdani}, which represent outputs as fuzzy sets requiring centroid-based defuzzification, TS systems assign numerical constants as rule consequents and compute the final output through a weighted average, which makes them faster and directly usable in control applications \cite{takagi1985sugeno}. Inference proceeds in four steps: each crisp input is fuzzified across all linguistic terms; each rule's firing strength is the product of its antecedent membership degrees (product t-norm); rules with non-zero firing strength are aggregated; and the output is defuzzified using:

\begin{equation}
y = \frac{\sum_{i=1}^{R} w_i \cdot c_i}{\sum_{i=1}^{R} w_i},
\label{eq:tsk-defuzz}
\end{equation}

\noindent where $w_i$ is the firing strength and $c_i$ is the consequent constant of rule $i$, summed over all $R$ fired rules. This computation completes in sub-millisecond time, adding minimal overhead at the 5-second monitoring interval.

\section{Experimental Methodology}
\label{sec:methodology}

Four experiments were conducted: (i) a multi-run variance experiment establishing reproducibility across fixed validator configurations, (ii) a unified seven-phase closed-loop experiment in which the controller triggers validator activation in response to a workload that grows and then declines, (iii) a comparative evaluation against three threshold-based controllers on the same unified workload, and (iv) an over-provisioned closed-loop experiment that exercises the \textit{Scale Down} direction.

\subsection{Testbed and Workload}
\label{subsec:testbed}

The Substrate network was configured with 10 genesis authorities, an AURA slot duration of 6 seconds, and GRANDPA finality. The number of active validators was varied across 4, 7, and 10 for the multi-run experiment, and started at 4 for the closed-loop experiments. When fewer than the full 10 nodes are active, AURA's round-robin assignment skips empty slots, increasing the effective block time. The expected effective block time for $n$ active nodes is $t_{\text{slot}} \times N / n$, where $N=10$ is the total authority count, giving 15.0 s, 8.57 s, and 6.0 s as nominal values for the three configurations. The fuzzy controller executes as a separate Python process that queries the blockchain via JSON-RPC every 5 seconds.

Real smart water meter data was sourced from the Queensland Government open data portal \cite{qld2022waterdata}, with daily Comma-Separated Values (CSV) files organised by meter count. Each file is SHA-256-hashed and the resulting 256-bit hash digest (H256, the Substrate hash type) is submitted on-chain through the custom pallet. Submission rate calibration established that the testbed sustains 75 extrinsics per second (ext/s) reliably. The unified workload therefore uses 75 ext/s as the heavy-load cap.

\subsection{Multi-run Variance Experiment}
\label{subsec:multi-run}

For each of the three configurations, the experiment was repeated five times. Each repeat begins with a clean chain state: chain data is purged, AURA (Sr25519) and GRANDPA (Ed25519) keys are inserted into each active node's keystore, and the network is launched. After stable block production is established, the load profile is applied while the controller samples the blockchain every 5 seconds for 420 seconds, recording all monitored variables at each interval. The five repeats support one-way Analysis of Variance (ANOVA) across configurations and pairwise Welch's t-tests on the pooled samples.

\subsection{Unified Closed-loop Experiment}
\label{subsec:dynamic}

\begin{table}[t]
\caption{Seven-Phase Unified Workload}
\label{tab:phases}
\centering
\footnotesize
\begin{tabular}{llll}
\toprule
\textbf{Phase} & \textbf{Time (s)} & \textbf{Load (ext/s)} & \textbf{Conditions} \\
\midrule
1. Idle          & 0-120     & 1            & Observing only \\
2. Ramp Up       & 120-300   & 5 $\to$ 75   & Triggers at 180 s \\
3. Heavy         & 300-540   & 75           & Sustained peak \\
4. Maintain High & 540-660   & 75           & Continued peak \\
5. Decline       & 660-840   & 75 $\to$ 2   & Load decreases \\
6. Light         & 840-1080  & 2            & Light load \\
7. Maintain Low  & 1080-1200 & 2            & Low equilibrium \\
\bottomrule
\end{tabular}
\end{table} 

The trigger logic is bidirectional by design, so the controller can both add and remove validators. \textit{Scale Up} fires when the action value crosses 0.7 in a single sample and fewer than 10 validators are active; \textit{Scale Down} fires when it falls below 0.3 and more than 4 validators are active. Whether a given direction is exercised depends on the workload: the cycle in Table \ref{tab:phases} is designed to drive the network through under-provisioned, target, and potentially over-provisioned states. After any scaling action, a 30-second stabilisation window suppresses further decisions. Node activation is verified through the peer count reported by the bootnode; deactivation is performed by sending SIGTERM to the highest-numbered active validator. At every monitoring sample, the controller records the firing strength of each of the 27 rules in addition to the aggregated output.

\subsection{Over-provisioned Closed-loop Experiment}
\label{subsec:scaledown-method}

To exercise the \textit{Scale Down} direction, a second closed-loop experiment starts from an over-provisioned configuration of eight active validators under light load. The controller, membership functions, rule base, and decision thresholds are identical to those above; only the initial validator count and the workload differ. The workload has three phases: an idle observation phase (0--120 s, 1 ext/s) followed by sustained light load (120--900 s, 2 ext/s). The initial count is eight rather than ten because the calibrated Node Count \textit{Many} term peaks at eight active validators, so starting at the chain ceiling of ten would place the input at the right foot of that term. An initial 60-second observation window precedes the first scaling decision, and the same 30-second stabilisation window applies after each action.

\subsection{Baselines and Statistical Analysis}
\label{subsec:baselines}

Three threshold-based controllers serve as baselines: Conservative (\textit{Scale Up} if block time $> 12$ s, $2.00\times$ slot), Moderate ($> 10$ s, $1.67\times$), and Aggressive ($> 8$ s, $1.33\times$). All three use a \textit{Scale Down} threshold of 7 s and a minimum of 4 active nodes. Each baseline runs the same seven-phase workload from identical starting conditions. Welch's t-tests are computed on block time samples aggregated into three operational regimes: \textit{Scale Up} (Phases 2-3), Maintain (Phases 4 and 7), and \textit{Scale Down} (Phases 5-6); Cohen's $d$ reports effect sizes. Decision flips, counted as transitions between adjacent recommendation categories in consecutive samples, measure controller stability.

\section{Results}
\label{sec:results}

This section reports the experimental results across the three studies described in Section \ref{sec:methodology}.

\subsection{Multi-run Variance Across Configurations}
\label{subsec:results-variance}

Table \ref{tab:variance} summarises the multi-run variance experiment. Each row reports the per-run mean and the within-run pooled standard deviation that captures the sample-level jitter.

\begin{table}[!htbp]
\caption{Multi-run Variance Across Configurations (5 Repeats Each)}
\label{tab:variance}
\centering
\footnotesize
\begin{tabular}{lccc}
\toprule
\textbf{Metric} & \textbf{4 active} & \textbf{7 active} & \textbf{10 active} \\
\midrule
Block time mean (s)            & 13.92 & 7.78  & 6.00 \\
Block time pooled std (s)      & 2.27  & 1.80  & 0.00 \\
Block size mean (MB)           & 0.029 & 0.024 & 0.019 \\
Efficiency mean (\%)           & 44.07 & 54.82 & 50.00 \\
Action value mean              & 0.639 & 0.435 & 0.500 \\
Blocks produced (mean)         & 27.2  & 48.6  & 69.4 \\
\bottomrule
\end{tabular}
\end{table}

The three configurations produce distinct operational profiles. The 10-node configuration maintains a constant 6.00-second block time, as the full set of authority slots is filled and no production opportunities are missed. The 7-node configuration averages 7.78 s; the effective block time formula gives $6 \times 10 / 7 \approx 8.57$ s, consistent with the measured value when adjacent filled slots partially offset the gaps. The 4-node configuration averages 13.92 s, close to $6 \times 10 / 4 = 15.0$ s within the within-run jitter.

The per-run means are deterministic given topology, because AURA's round-robin assignment is fixed once the active set is decided, giving a between-run standard deviation of approximately zero. The meaningful variance lies within each run, as the rotation passes through filled and empty slot sequences; this is captured by the within-run pooled standard deviation.

Because the per-run means are deterministic, the run-level statistics are highly concentrated: treating each run's mean as a single observation, the between-run standard deviation is effectively zero (below 0.01 s) for all three configurations, so the run-level 95\% confidence intervals on mean block time are narrower than $\pm 0.01$ s. The configuration means (13.92, 7.78, and 6.00 s) are separated by far more than this spread, so the differences remain significant even when only the five run-level means per configuration are treated as independent observations. This addresses the concern that pooled time-series samples within a run may be autocorrelated.

\textbf{Statistical tests.} One-way ANOVA across configurations on the pooled samples produces $F=7229$ for block time, $F=1042$ for efficiency, and $F=2215$ for action value, all with $p<0.001$. Pairwise Welch's t-tests on the within-run pooled samples confirm that all configuration pairs differ significantly ($p<0.001$) on every measured metric. The 4-vs-10 comparison on block time produces $t=71.37$, the largest magnitude in the analysis.

\subsection{Unified Closed-loop Experiment}
\label{subsec:results-dynamic}

\begin{figure*}[!htbp]
    \centerline{\includegraphics[width=0.68\textwidth]{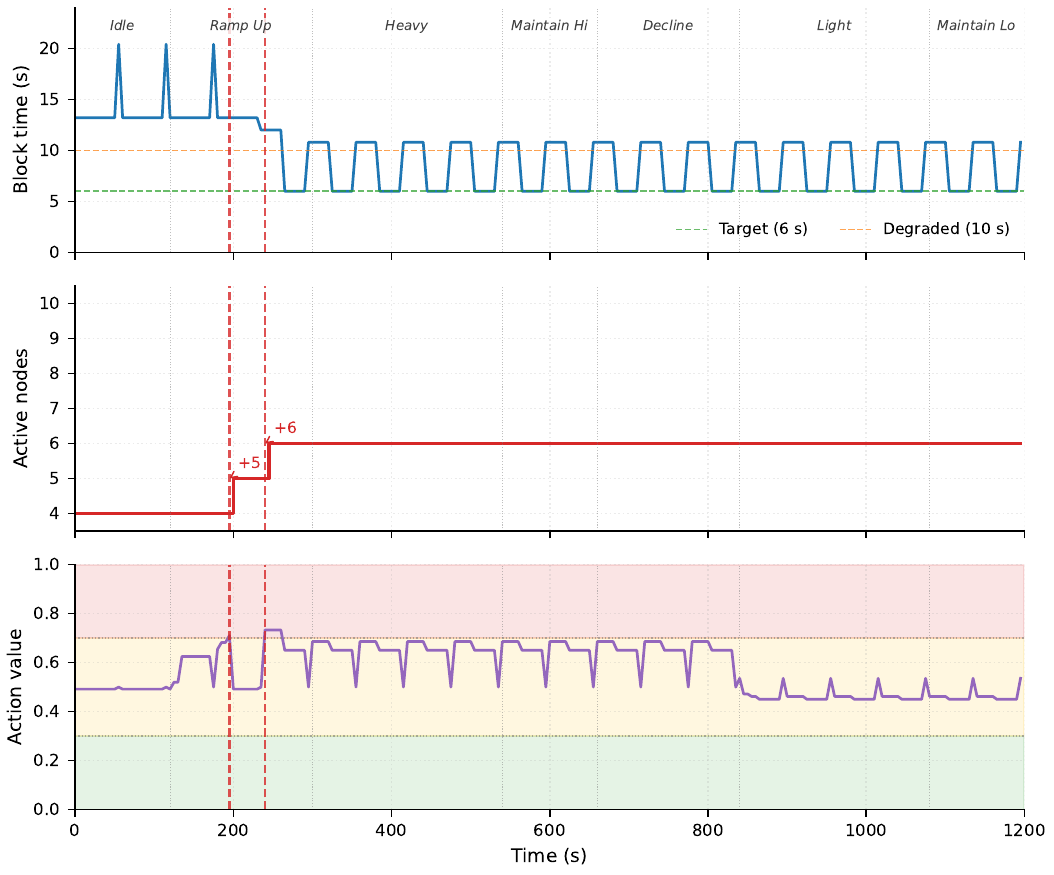}}
    \caption{Unified closed-loop experiment, 1200 s, seven phases. Top: block production time, with horizontal lines at 6 s (target) and 10 s (degraded). Middle: active node count, showing the two \textit{Scale Up} events (4$\to$5 at $t=195$ s, 5$\to$6 at $t=240$ s). Bottom: action value with decision zone shading (green: \textit{Scale Down}; yellow: \textit{Maintain}; red: \textit{Scale Up}). }
    \label{fig:dynamic}
\end{figure*}

Fig. \ref{fig:dynamic} presents the time series of the 1200-second unified closed-loop experiment across all seven phases.

The controller triggers two \textit{Scale Up} events during Phase 2. At $t=195$ s the action value crosses 0.7 and the controller activates a fifth validator (peer discovery in 2.04 s); block time drops from 14.4 s to 10.2 s after the 30-second stabilisation window, within 1.8 s of the new effective slot target of 12.0 s. At $t=240$ s a sixth validator is activated; block time drops from 13.0 s to 8.4 s, within 1.6 s of the 10.0 s target for six active nodes.

After $t=240$ s the controller holds the six-node configuration for the remaining 960 s. Block time settles at 8.4 s, slightly below the 10.0 s target, indicating that six validators are sufficient for the sustained load through Phases 3, 4, and 5. The controller produces no further scaling actions through Phases 6 or 7. The per-phase recommendation distributions are: 100\% \textit{Maintain} in Phases 1, 3, 4, 5, 6, and 7; 83.3\% \textit{Maintain} and 16.7\% \textit{Scale Up} in Phase 2.

These distributions support H1a (two \textit{Scale Up} events during the load increase when block time was elevated relative to the four-node target) and H1b (100\% \textit{Maintain} in both six-node equilibria, Phases 4 and 7). H1c (a \textit{Scale Down} event under over-provisioning) was not triggered by this workload, which never drove the network above its target provisioning; it is demonstrated separately in Section \ref{subsec:results-scaledown}. H2 is supported by both scaling events: post-stabilisation block times approached the new effective slot target, reducing block time by 4.2 s and 4.6 s from pre-event values.

\textbf{Rule firing coverage:} Six of the 27 rules produced firing strength of at least 0.1 in this experiment: R11, R14, R17, R20, R23, R26. R20 (Medium BT, Small BS, Moderate NC) dominates the Idle phase; R17 and R26 drive the Heavy and \textit{Maintain High} phases as block size grows under the 75 ext/s load; R11 takes over when block size returns to baseline. The over-provisioned experiment (Section \ref{subsec:results-scaledown}) activates a seventh rule, R12, raising the combined coverage to seven rules across the two closed-loop experiments. The rules that remain unfired correspond to input combinations this architecture cannot produce: because block production time in an AURA chain is governed by the active validator count rather than by transaction volume, states such as High block time with Many active validators, or High block time with Small block size, do not co-occur in this testbed. These rules are retained for completeness and for deployments with different consensus or storage characteristics, consistent with the anchoring framework of Section \ref{subsec:mf-calibration}.

\subsection{Comparison Against Threshold Baselines}
\label{subsec:results-comparison}

Table \ref{tab:comparison} compares the four controllers on the identical 1200-second unified workload. Each controller starts at four active validators with the same load profile.

\begin{table}[t]
\caption{Per-controller Comparison Metrics (Identical Unified Workload)}
\label{tab:comparison}
\centering
\footnotesize
\begin{tabular}{lccccc}
\toprule
\textbf{Controller} & \textbf{Up} & \textbf{Down} & \textbf{Flips} & \textbf{Final BT (s)} & \textbf{Final Eff (\%)} \\
\midrule
TS Fuzzy        & 2  & 0  & 4  & 8.4  & 67.9 \\
Conservative    & 3  & 1  & 4  & 12.0 & 40.0 \\
Moderate        & 18 & 14 & 29 & 7.9  & 81.0 \\
Aggressive      & 17 & 12 & 24 & 11.0 & 50.0 \\
\bottomrule
\end{tabular}
\end{table}

The four controllers differ markedly. The TS controller issues 2 scaling actions and 4 decision flips over 1200 s. The Conservative produces 4 flips but settles at a final efficiency of only 40\%. Moderate and Aggressive are highly reactive, producing 24-29 flips across 17-18 \textit{Scale Up} and 12-14 \textit{Scale Down} actions.

\begin{figure*}[!htbp]
    \centerline{\includegraphics[width=0.68\textwidth]{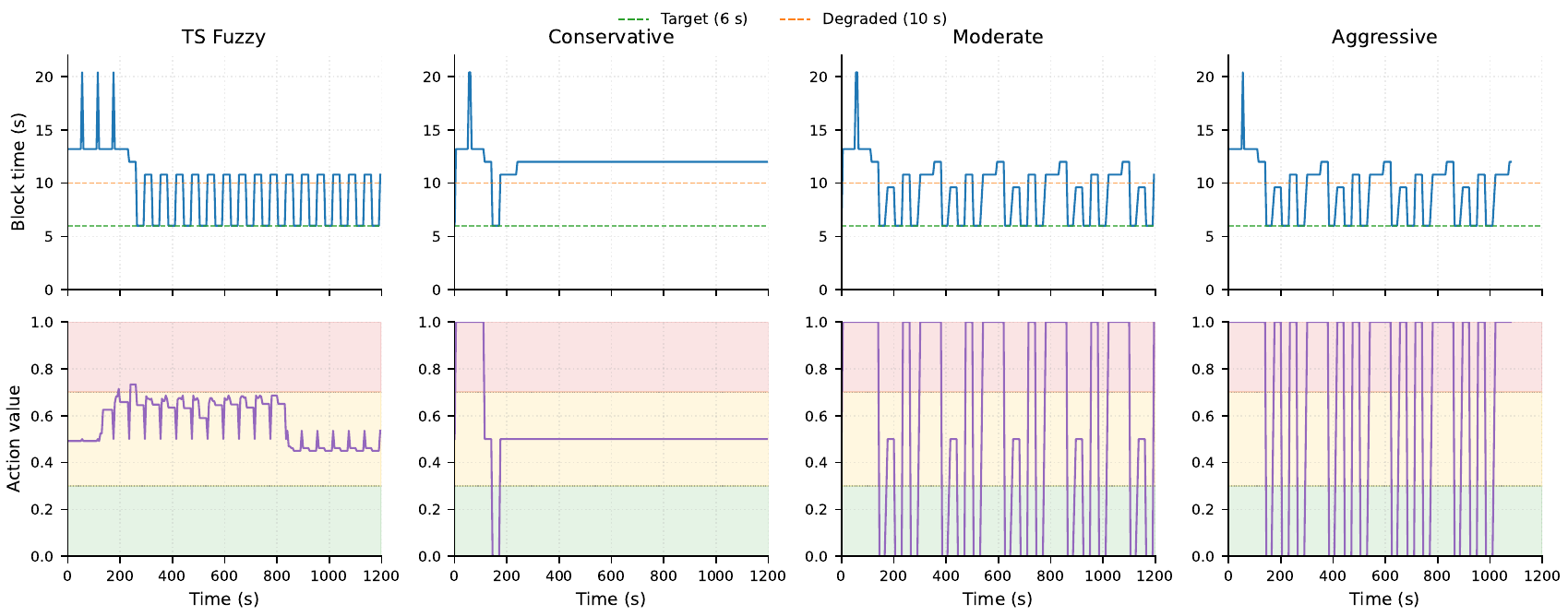}}
    \caption{Four-controller comparison under the identical unified workload. Top row: block production time with reference lines at 6 s (target) and 10 s (degraded). Bottom row: action value with decision zone shading. The TS controller produces a smooth progression; the threshold controllers oscillate around their fixed cut-offs.}
    \label{fig:comparison}
\end{figure*}

Fig. \ref{fig:comparison} visualises the contrast. The TS column shows a smooth action value progression that remains predominantly in the Maintain zone. Moderate and Aggressive show pronounced square-wave oscillation as block time crosses and re-crosses the fixed thresholds; Conservative oscillates less but at a much higher block time band.

\textbf{Statistical comparison:} Table \ref{tab:ttest} reports pairwise Welch's t-tests on block time aggregated into three operational regimes. Against Conservative, the TS controller produces significantly lower block times across all three regimes, with large effect sizes ($d = -2.10$ and $-2.11$ in Maintain and \textit{Scale Down}). Against Moderate, the TS controller is slightly slower in the \textit{Scale Up} regime (the threshold scales earlier and more often) but differences in \textit{Maintain} and \textit{Scale Down} are small and not significant. Against Aggressive, a similar pattern holds with a medium effect in Maintain ($d = -0.73$).

The decision flip counts in Table \ref{tab:comparison} are the strongest discriminator: 4 flips for TS against 24 and 29 for the reactive thresholds. This is the core stability-reactivity trade-off: more reactive controllers occasionally achieve lower block times at the cost of substantially more scaling churn, while the TS controller's smoother decision surface produces a quiet operating regime with comparable end-state performance.

\begin{table}[t]
\caption{Welch's t-tests on Block Time, TS vs Threshold (by Regime)}
\label{tab:ttest}
\centering
\footnotesize
\setlength{\tabcolsep}{4pt}
\begin{tabular}{llccc}
\toprule
\textbf{TS vs} & \textbf{Regime} & \textbf{$t$} & \textbf{$p$} & \textbf{$d$} \\
\midrule
Conservative & Scale Up    & $-3.78$  & $<0.001$ & $-0.58$ \\
Conservative & Maintain    & $-10.28$ & $<0.001$ & $-2.10$ \\
Conservative & Scale Down  & $-13.67$ & $<0.001$ & $-2.11$ \\
\midrule
Moderate     & Scale Up    & $\phantom{-}2.56$    & $0.012$   & $\phantom{-}0.39$  \\
Moderate     & Maintain    & $-1.79$  & $0.077$   & $-0.36$ \\
Moderate     & Scale Down  & $-1.84$  & $0.068$   & $-0.28$ \\
\midrule
Aggressive   & Scale Up    & $\phantom{-}2.56$    & $0.012$   & $\phantom{-}0.39$  \\
Aggressive   & Maintain    & $-3.11$  & $0.003$   & $-0.73$ \\
Aggressive   & Scale Down  & $-1.84$  & $0.068$   & $-0.28$ \\
\bottomrule
\end{tabular}
\vspace{1pt}\\
{\footnotesize $d$ is Cohen's effect size. Negative $d$: TS block time lower than baseline.}
\end{table}

\subsection{Scale Down Under Over-Provisioning}
\label{subsec:results-scaledown}

\begin{figure}[t]
    \centerline{\includegraphics[width=\columnwidth]{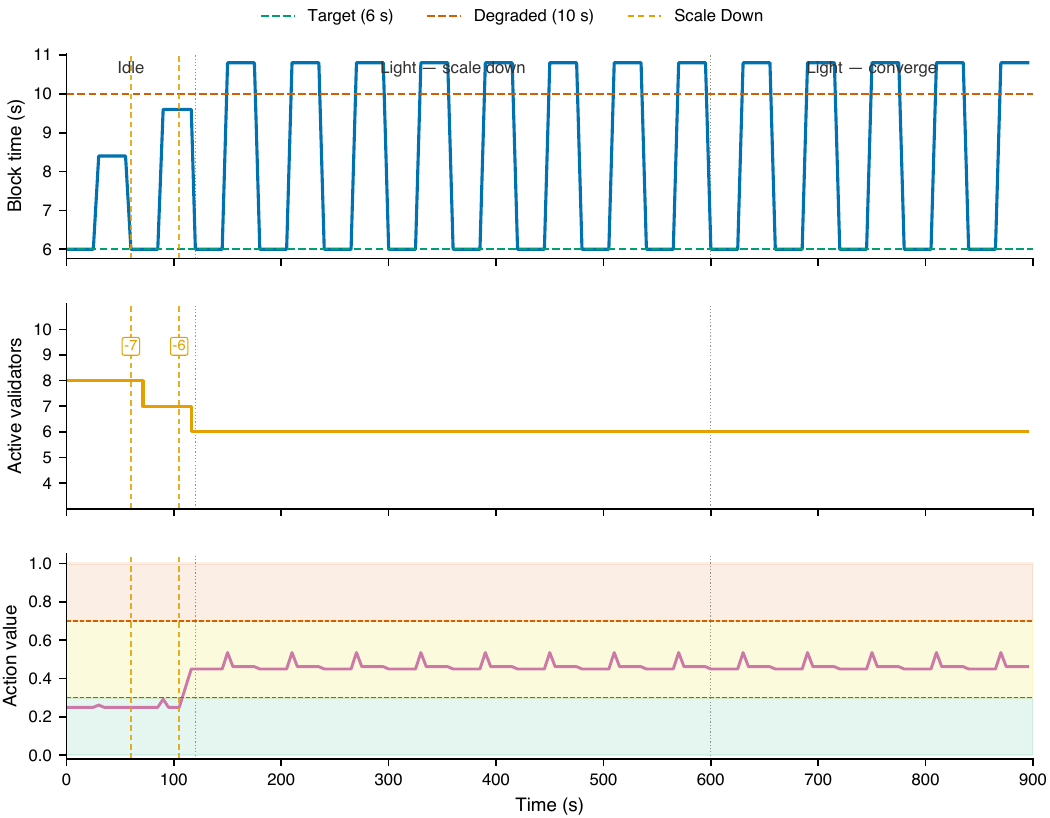}}
    \caption{Over-provisioned closed-loop experiment, 900 s, starting from eight active validators under light load. Top: block production time, with reference lines at 6 s (target) and 10 s (degraded). Middle: active node count, showing two \textit{Scale Down} events (8$\to$7 at $t=60$ s, 7$\to$6 at $t=105$ s). Bottom: action value with decision zone shading (green: \textit{Scale Down}; yellow: \textit{Maintain}; red: \textit{Scale Up}). The controller converges to the same six-validator equilibrium reached from below in Fig. \ref{fig:dynamic}.}
    \label{fig:scaledown}
\end{figure}

The unified workload exercised \textit{Scale Up} and \textit{Maintain}. The over-provisioned experiment (Section \ref{subsec:scaledown-method}) exercises the \textit{Scale Down} direction. Fig. \ref{fig:scaledown} presents the 900-second time series, starting from eight active validators under light load.

At eight active validators the action value held at 0.250, in the \textit{Scale Down} region, with block time oscillating between 6.0 and 8.4 s as AURA rotates through filled and empty slots. The controller deactivated one validator at $t=60$ s (eight to seven active) and a second at $t=105$ s (seven to six active), both during the idle phase, each followed by the 30-second stabilisation window. After convergence, block time at six active validators averaged 8.4 s (oscillating between 6.0 and 10.8 s) and the action value rose to a mean of 0.46, within the \textit{Maintain} band. The controller held the six-validator configuration for the remainder of the run, with no further scaling. The recommendation distribution was 91.7\% \textit{Scale Down} during the idle phase and 100\% \textit{Maintain} thereafter.

This confirms convergence from above: the over-provisioned eight-validator network was reduced to the same six-validator equilibrium reached from below in the unified experiment, at the same mean block time of 8.4 s. The over-provisioning rule R12 (Medium BT, Small BS, Many NC) fired only in this experiment, dominating the idle phase before convergence; at equilibrium R11 again dominates. Across both closed-loop experiments the controller exercised both scaling directions, each converging to a stable equilibrium.

\section{Discussion}
\label{sec:discussion}

\subsection{Stability versus Reactivity}
\label{subsec:stability}

The most prominent pattern is the trade-off between scaling stability and reactivity. The TS controller produces 4 decision flips against 24 to 29 for the Moderate and Aggressive variants, an order-of-magnitude difference. The threshold logic produces this oscillation directly: a scaling decision changes category each time block time crosses the fixed cut-off. Because block time naturally fluctuates around the effective slot duration as AURA rotates through validators, the threshold controllers spend much of the experiment in flux.

The TS controller avoids this through degrees of membership: a block time of 9.6 s holds partial membership in both Medium and High, so firing strengths combine smoothly and the output changes gradually with the input rather than discontinuously. For production deployments where each scaling action involves keystore configuration, peer discovery, and chain synchronisation, two actions are far preferable to seventeen. Notably, the proposed controller already incorporates a cooldown policy in the form of the 30-second stabilisation window, so the baselines and the TS controller are held to the same stabilisation constraint; the difference in stability therefore stems from fuzzy inference rather than from cooldown alone.

\subsection{Generalisability and Determinism}
\label{subsec:generalisability}

The empirical recalibration framework in Section \ref{subsec:mf-calibration} addresses a practical concern with fuzzy controllers, namely that MF parameters are often tied  to specific deployment characteristics. The refinement shifts the BT High triangle peak from 20 s (the worst-case prediction from the effective slot formula) to 14 s (the actual observed worst case), and moves the NC Many peak from 10 (the chain authority ceiling) to 8 (the realistic ``fully scaled'' point for a controller starting at four active validators). The worked examples in Table \ref{tab:mf-examples} show that the same anchoring rule transfers across chains with substantially different slot durations. The 27 rules and their consequents transfer without modification.

\section{Conclusion}
\label{sec:conclusion}

This paper presented a closed-loop dynamic validator node scaling system for private Substrate blockchains, with a TS fuzzy inference controller that monitors live blockchain parameters and triggers validator activation through verified peer discovery. The MFs are anchored to the empirically observed operating range, with a generalisable anchoring rule and four worked examples covering different deployment configurations.

The experimental evaluation tested three hypotheses on a 10-authority Substrate testbed with real smart water meter data. The controller produced statistically distinct operational profiles across configurations of 4, 7, and 10 active validators (one-way ANOVA $F=7229$ on block time, $p<0.001$). In a unified seven-phase 1200-second closed-loop experiment, the controller triggered two \textit{Scale Up} events as load increased, reaching a six-validator equilibrium that held through both the sustained peak load and the return to baseline. A separate over-provisioned experiment exercised the \textit{Scale Down} direction, converging from eight validators to the same six-validator equilibrium. Against three threshold baselines, the TS controller produced 4 decision flips compared with 24 to 29 for the more reactive thresholds while maintaining comparable block times, with Welch's t-tests confirming large effect sizes ($|d|$ up to $2.11$). Future work will develop an evolving variant that adapts MF parameters online, compare against hysteresis and PID baselines, and add complementary fuzzy subsystems for storage configuration and archival management.

\section*{Acknowledgment}
A use of generative AI assistant (Anthropic Claude) for language editing. All technical content, including the system design, experiments, analyses, and conclusions, is the authors' own, and the authors reviewed and verified all text in the paper.  

\bibliographystyle{IEEEtran}
\bibliography{references}

\end{document}